\documentclass[a4paper,12pt]{amsart}

\usepackage{amsmath}
\usepackage{amssymb}
\usepackage{amsthm}
\usepackage{hyperref}
\usepackage{cite}

\newcommand{\fU}{\mathfrak{U}}
\newcommand{\fV}{\mathfrak{V}}
\newcommand{\bD}{\boldsymbol{\Delta}}
\newcommand{\bU}{\mathbf{U}}
\newcommand{\bV}{\mathbf{V}}

\newcommand{\bZ}{\mathbf{Z}}
\newcommand{\Pp}{\mathcal{P}}
\newcommand{\wtw}{\widetilde{w}}
\newcommand{\wtl}{\widetilde{\lambda}}
\newcommand{\wtO}{\widetilde{\Omega}}
\newcommand{\wto}{\widetilde{\omega}}
\newcommand{\wtP}{\widetilde{\Pp}}
\newcommand{\wtK}{\widetilde{K}}
\newcommand{\wtV}{\widetilde{\bV}}
\newcommand{\wtPsi}{\widetilde{\Psi}}
\newcommand{\wtpsi}{\widetilde{\psi}}
\newcommand{\wtU}{\widetilde{\bU}}
\newcommand{\wtPhi}{\widetilde{\Phi}}

\newtheorem{thm}{Theorem}

\allowdisplaybreaks

\begin{document}

\title[Some comments \ldots]{Some  comments on  continuous symmetries  of AKNS  hierarchy equations and their solutions}

\author{A. O.~Smirnov}
\email{\href{mailto:alsmir@guap.ru}{alsmir@guap.ru}}

\address{Sankt-Petersburg State University of Aerospace  Instrumentation, Sankt-Petersburg , Russia}

\author{V.B.~Matveev}
\email{\href{mailto:vladimir.matveev@u-bourgogne.fr}{vladimir.matveev@u-bourgogne.fr}}

\address{Sankt-Petersburg State University of Aerospace  Instrumentation, Sankt-Petersburg , Russia}

\address{Institut   de Math\'ematiques de Bourgogne (IMB), Universit\'e de Bourgogne-Franche Comt\'e, Dijon, France}

\maketitle
\begin{abstract}
 We show that  a well known for NLS equation  scaling invariance and Galilean invariance property of its solutions
can be extended
with appropriate modifications  to the whole AKNS hierahchy and its reduced and deformed versions.

\end{abstract}



\section{Introduction: Integrable systems described by AKNS hierarchy}

 The k-th equation of the AKNS hierarchy results from the compatibility condition
  \begin{equation}
\label{eq:c}
 (\Psi_x)_{t_k}=(\Psi_{t_k})_x
\end{equation}
 of the following Lax system:
\begin{equation}
\begin{cases}
\Psi_x=\fU \Psi,\\
\Psi_{t_k}=\fV_k \Psi,
\end{cases}
\label{eq:Lax}
\end{equation}
\begin{gather*}
\fU:=\lambda J+\fU^0,\quad \fV_1:=2\lambda \fU+\fV_{1}^0,\quad \fV_{k+1}:=2\lambda \fV_k+\fV_{k+1}^0,\quad k\geqslant1\\
J:=\begin{pmatrix} -i&0\\0&i \end{pmatrix}, \quad
\fU^0:=\begin{pmatrix} 0&i\psi\\ -i\phi&0 \end{pmatrix}.
\end{gather*}
Equation
 \eqref{eq:c} determines  the form of the matrices  $\fV_k$, the recursion relations between the  off-diagonal  elements of   the matrices  $\fV_k^0$, and the relations between  the diagonal and  off-diagonal  elements of
these matrices:
\begin{equation*}
[J,\fV_1^0]=2(\fU^0)_x,\quad [J,\fV_{k+1}^0]=2(\fV_k^0)_x+2[\fV_k^0,\fU^0],\quad k\geqslant 1.
\end{equation*}
In particular,
\begin{equation*}
\fV_1^0=\begin{pmatrix}-i\psi\phi& -\psi_x\\-\phi_x &i\psi\phi\end{pmatrix},\quad
\fV_2^0=\begin{pmatrix} \psi_x\phi-\phi_x\psi& 2i\psi^2\phi-i\psi_{xx}\\
-2i\phi^2\psi+i\phi_{xx}& \phi_x\psi-\psi_x\phi\end{pmatrix}.
\end{equation*}
The evolution equation, describing the dynamics  of the matrix elements of  $\fU^0$:
\begin{equation}
(\fU^0)_{t_k}=(\fV_k^0)_x+[\fV_k^0,\fU^0]=\dfrac12[J,\fV_{k+1}^0],
\end{equation}
also follows from \eqref{eq:c}. It is called the k-th member of the AKNS hierarchy.
The complementary restriction  $\phi=-\psi^\ast$ transform the AKNS hierarhy to the reduced AKNS
hierarchy (RAKNS hierarchy).  Here we mainly restrict our discussion by considering scalar nonlinear PDE's belonging to RAKNS hierarchy although all our results can be easily extended to the AKNS case.
First member of the AKNS hierarchy  is the coupled NLS system:
\begin{equation*}
\begin{cases}
i\psi_{t_1}+\psi_{xx}-2\psi^2\phi=0,\\
-i\phi_{t_1}+\psi_{xx}-2\phi^2\psi=0.
\end{cases} \label{eq:coup.nls}
\end{equation*}
 The first member of the RAKNS hierachy is just a focusing NLS equation:
\begin{equation}
i\psi_{t_1}+\psi_{xx}+2|\psi|^2\psi=0. \label{eq:nls}
\end{equation}

Second member of AKNS  hierarchy is a  modified KdV (MKdV) system:
\begin{equation*}
\begin{cases}
\psi_{t_2}+\psi_{xxx}-6\psi\phi\psi_x=0,\\
\phi_{t_2}+\phi_{xxx}-6\psi\phi\phi_x=0.
\end{cases} \label{eq:coup.mkdv}
\end{equation*}
Its RAKNS counterpart  is the scalar (complex) MKdV equation\footnote{Recall that the usual mKdV
reads $\psi_{t_2}+\psi_{xxx}+6\psi^2\psi_x=0,$ and only its real solutions are of physical interest.}
\begin{equation}
\psi_{t_2}+\psi_{xxx}+6|\psi|^2\psi_x=0. \label{eq:mkdv}
\end{equation}

Third member of AKNS hierarchy is a following coupled system of nonlinear PDE's:
\begin{equation*}
\begin{cases}
i\psi_{t_3}-\psi_{xxxx}+8\psi\phi\psi_{xx}+2\psi^2\phi_{xx}+6\psi_x^2\phi+4\psi\psi_x\phi_x-6\psi^3\phi^2=0,\\
-i\phi_{t_3}-\phi_{xxxx}+8\psi\phi\phi_{xx}+2\phi^2\psi_{xx}+6\psi\phi_x^2+4\phi\psi_x\phi_x-6\psi^2\phi^3=0.
\end{cases} \label{eq:coup.akns3}
\end{equation*}
Its RAKNS counterpart  for  $t_3=-t$ becomes a well known  Lakshmanan-Porcezian-Daniel equation
(LPD equation)  \cite{GNLS88, GNLS92, GNLS93}:
\begin{equation} \label{eq:lpd}
i\psi_{t}+\psi_{xxxx}+8|\psi|^2\psi_{xx}+2\psi^2\psi^{\ast}_{xx}+6\psi_x^2\psi^{\ast}+4\psi|\psi_x|^2+6|\psi|^4\psi=0.
\end{equation}

Higher order equations are rarely considered by physicists. For completness we  display here  also  a 4-th  member of RAKNS hierarchy (RAKNS-4 equation) for $t_4=-t$:\footnote{Surprisingly it contains less terms with respect to \eqref{eq:lpd}.}
\begin{equation} \label{eq:akns4}
\psi_{t}+\psi_{5x}+10|\psi|^2\psi_{xxx}+20\psi_{xx}\psi_x\psi^\ast+10(|\psi_x|^2\psi)_x+30|\psi|^4\psi_x=0,
\end{equation}
and the 5-th member of RAKNS hierarchy (RAKNS-5 equation) with $t_5= t$:
\begin{multline} \label{eq:akns5}
i\psi_{t}+\psi_{6x}+12|\psi|^2\psi_{xxxx}+2\psi^2\psi^\ast_{xxxx}+30\psi_{xxx}\psi_x\psi^\ast
+18\psi_{xxx}\psi\psi^\ast_x+8\psi_x\psi\psi^\ast_{xxx}+\\
+50\psi_{xx}|\psi_x|^2+50\psi_{xx}|\psi|^4+20\psi_{xx}^2\psi^\ast+22|\psi_{xx}|^2\psi+20\psi_x^2\psi^\ast_{xx}
+20|\psi|^2\psi^2\psi^\ast_{xx}+\\
+10\psi^3(\psi^\ast_x)^2+70\psi_x^2|\psi|^2\psi^\ast+60|\psi|^2|\psi_x|^2\psi+20|\psi|^6\psi =0.
\end{multline}

Below we  use for all equations of the RAKNS hierarchy the following  shortened notations:
\begin{gather*}
i\psi_{t_1}+H_1(\psi)=0,\quad
\psi_{t_2}+H_2(\psi)=0,\quad
-i\psi_{t_3}+H_3(\psi)=0,\\
-\psi_{t_4}+H_4(\psi)=0,\quad
i\psi_{t_5}+H_5(\psi)=0,
\end{gather*}
etc or equivalently
\begin{equation*}
\psi_{t_k}=i^{k+1}H_k(\psi).
\end{equation*}

The key property  of the considered equations   is the existence of the joint solutions
\begin{equation*}
\Psi(x,t_1,\ldots,t_k,\ldots ),
\end{equation*}
satisfying all the  AKNS equations simultaneously. When all times  except one selected $t_j$  are fixed this is still the solution to the j-th equation of the hierarchy where the other times play the role of parameters. Formal solutions of this kind are provided by the Sato theory  and include the infinite number of times.
See for instance \cite{Ohta} for an elementary introduction to the Sato theory.

We restrict our consideration by the case of any finite number of times. In this case the related common solutions are the finite-gap solutions  of the fixed genus or their degenerate
cases which are very diversified \cite{BBEIM}.  If the order of the considered equation is less than a number of phases of the solution
variables belonging to the higher order phases plays a role of free parameters of the solution. It is worthwhile to mention that when  the hierarchic  order of equation $n$  is greater than a number of phases  $k$  then $\Psi$-function  either is independent from  $t_n$,   or $t_n $ enters into the lower order phases $ T_j$   in a linear way.

\section{Mixed RAKNS equations}

Together  with individual members of  AKNS  hierarchy the mixed equations, corresponding to some  selection of finite number  of times which are identified, and denoted by the single letter $t$ are also  frequently  considered, obviously leading also to the integrable nonlinear PDE's having in their RHS  the linear combinations of the RHS of the arbitrary  chosen  individual members of the hierarchy.

The most  popular  example of this kind is the Hirota integrable equation  \cite{Hir73, Hir06, AAS2, Hir13, HeLiP}:
\begin{equation}
i\psi_t+\alpha H_1(\psi)-i\beta H_2(\psi)=0,  \quad  \alpha, \beta \in \mathbb{R}.
\label{eq:hir}
\end{equation}
Obviously  this equation  admits the solutions of the form  $\Psi(x,\alpha t,-\beta t,\ldots,t_k)$,
where  $\Psi(x,t_1,\ldots,t_k)$ -- is an arbitrary solution  of the AKNS hierarchy.
More complicated  "mixed"  equation  \cite{GNLS13, AA14}, involving a linear combination of first 3 members of the  RAKNS hierarchy is
\begin{equation}
i\psi_t+\alpha H_1(\psi)-i\beta H_2(\psi)+\gamma_1 H_3(\psi)=0. \label{eq:gnls}
\end{equation}
Its solutions are given by  $\Psi(x,\alpha t,-\beta t,-\gamma_1 t,\ldots,t_k)$.
Next by the level of complexity mixed models  are  described by the following PDE's:
\begin{equation}
i\psi_t+\alpha H_1(\psi)-i\beta H_2(\psi)+\gamma_1 H_3(\psi)-i\gamma_2 H_4(\psi)=0,
\label{eq:hnls4}
\end{equation}
and
\begin{equation}
i\psi_t+\alpha H_1(\psi)-i\beta H_2(\psi)+\gamma_1 H_3(\psi)-i\gamma_2 H_4(\psi)+\gamma_3 H_5(\psi)=0,
\label{eq:hnls5}
\end{equation}
A particular case of \eqref{eq:hnls4} with  $\beta=0$, $\gamma_1=0$ was considered very recently in in \cite{AKAC}.
In particular,  in \cite{AKAC}   were studied the 2-breathers solutions  and its limit case -- quasi-rational MRW
solution of rank 2.

More generally, it  is a well known fact that  all equations of the form
\begin{equation}
\psi_t  = \sum_{k \geq 1}^{N}  i^{k}b_{k}H_k(\psi),\quad   b_k \in \mathbb{R}.
\label{eq:Mix}
\end{equation}
are completely integrable  and their  stationary solutions are the algebro-geometric finite gap potentials expressed by means of Riemann theta functions of the hyperelliptic curves  or their degenerate cases.

\section{Deformed  RAKNS hierarchy}
Suppose that  $t_k=\alpha_{k}(t) $ in AKNS or  RAKNS hierarchy  where $\alpha_{k}(t)$  are any (in general different for different $k$-s)  differentiable functions.
It is obvious that the change of variable $t_j = \alpha_{j}(t)$ transforms the second  of equation \eqref{eq:Lax} into the following linear equation for the matrix function
$\Psi$ :
\begin{equation}
\label{eq:t}
\Psi_t= \alpha_j'(t)V_j \Psi.
\end{equation}.
The compatibility  condition of \eqref{eq:t}   with first  equation of the system \eqref{eq:Lax} now takes the form:
\begin{equation}
U_t + \alpha'_{j}(t)\left( - (V_j )_x + [ U,V_j] \right) =0,
\label{eq:L2}
\end{equation}
and the j-th equation of thus obtained hierarchy, which we call deformed  RAKNS hierarchy,  reads:
\begin{equation}
\psi_t =i^{k}\alpha'_{j}(t)H_j
\label{eq:def}
\end{equation}
Now it is clear that the function
 \begin{equation}
\label{eq:Sol}
\Psi(x,\alpha_1(t),\ldots,\alpha_k(t))
\end{equation}
 furnishes  the solutions of  the nonlinear PDE
\begin{equation}
i\psi_t-\sum_{k\ge1}i^{k+1}\alpha'_k(t)H_k=0,
 \label{eq:tnls}
\end{equation}
which contains   the "mixed" RAKNS equations obtained from the linear combinations of the functions
of the RHS of  RAKNS equations with constant coefficients $b_{k}$, as a special case  where  $\alpha_{j}(t) := b_{k}t  + d_k$, and  $b_k, d_k$ are any different real constants.\footnote{The same  equations written in a  form  slightly different from our presentation  were first mentioned  in
 \cite{AKAC15}  (to appear    in Chaos).
 }
The simple explanation of their nature given here,  simply coming from the general change of time variables in (the system ) \eqref{eq:Lax}  seems to be  more  transparent with respect to
\cite{AKAC15}.  Our  logic immediately furnish the large variety of solutions from generic  solutions of the RAKNS hierarchy.
It is also quite obvious that choosing $\alpha_{j}(t) $ with nonintersecting compact supports $\Delta_j$  we obtain the solutions of \eqref{eq:tnls}, which,   when $t \in \Delta_j $,  is the same as the solution
the j-th equation of the deformed hierarchy:
$$ \psi_t = i^{k}\alpha_{j}'(t) H_{j}(\psi). $$

\section{Continuous  symmetries  of the RAKNS hierarchy}
All   RAKNS hierarchy equations are invariant both with respect to the space and time translations  and  with respect  to the constant phase transformations $\psi(x,t) \to e^{i\varphi}\psi(x,t), \varphi \in \mathbb{R}$. Less trivial are the scaling invariance  and
the Galilean invariance properties  depending  on the chosen equation of hierarchy or their mixture.
For instance,  the scaling covariance  property of the n-th member of the  RAKNS hierarchy takes the form:
 $$ \psi_{n}(x,t)  \to q\psi(qx,q^{n+1}t), \quad q>0, $$
which means that the transformation above maps the solution $\psi(x,t) $ to the new solution of the same equation. In particular this means that the study of the solutions, having  asymptotic magnitude $q $
when $x^2 + t^2 \to \infty $ , can always be  be reduced to the case $q=1$.

The proposition formulated below extends the scaling invariance and Galilean invariance properties to the
the whole RAKNS hierarchy and its mixed and times deformed versions discussed above.

Suppose  $a,b\in\mathbb{R}$ and
\begin{align*}
&X:=ax+a\sum_{m\ge1}\binom{m+1}{1}(2b)^mt_m,\\
&T_1:=a^2t_1+a^2\sum_{m\ge2}\binom{m+1}{2}(2b)^{m-1}t_m,\\
&T_2:=a^3t_2+a^3\sum_{m\ge3}\binom{m+1}{3}(2b)^{m-2}t_m,\\
&\quad\ldots
\end{align*}

\begin{thm}
Suppose the function
\begin{equation*}
\Psi(x,t_{1},t_{2},\ldots)
\end{equation*}
is a solution of the AKNS hierarchy equations. Then
\begin{equation*}
\wtPsi(x,t_1,t_2,\ldots)= a\Psi\left(X,T_1,T_2,\ldots\right)\exp\left\{-2ibx-i\sum_{m\ge1}(2b)^{m+1}t_m\right\}
\end{equation*}
a new solution of the same hierarchy.
\end{thm}
Suppose now that $b_k,a,b\in\mathbb{R}$,
\begin{align*}
&X:=ax+at\sum_{m\ge1}\binom{m+1}{1}(2b)^mb_m,\\
&T_1:=\left(b_1+\sum_{m\ge2}\binom{m+1}{2}(2b)^{m-1}b_m\right)a^2t,\\
&T_2:=\left(b_2+\sum_{m\ge3}\binom{m+1}{3}(2b)^{m-2}b_m\right)a^3t,\\
&\quad\ldots
\end{align*}

\begin{thm}
Suppose that the function
\begin{equation*}
\Psi(x,b_{1}t,b_{2}t,\ldots)
\end{equation*}
satisfies the mixed RAKNS equation
\begin{equation}
i\psi_t-\sum_{m\ge1}i^{k+1}b_kH_k(\psi)=0.
\label{gnls}
\end{equation}
Then the function
\begin{equation*}
\wtPsi(x,t)= a\Psi\left(X,T_1,T_2,\ldots\right)\exp\left\{-2ibx-it\sum_{m\ge1}(2b)^{m+1}b_m\right\}
\end{equation*}
is also a solution of  \eqref{eq:gnls}.

\end{thm}

In particular, assuming that  $\psi(x,\alpha t, -\beta t)$ is a solution of the Hirota equation  \eqref{eq:hir},
the function
\begin{equation*}
\wtpsi=a\psi(ax+4[\alpha-3b\beta]abt,[\alpha-6b\beta]a^2t,-\beta a^3t)\exp\{-2ibx-4i(\alpha-2\beta b)b^2t\}.
\end{equation*}
is again the  solution of the same equation.

The transformation $\psi \to \wtpsi $
\begin{equation*}
\wtpsi(x,t)=a\psi(ax+4abt,a^2t)\exp\{-2ibx-4ib^2t\}.
\end{equation*}
in this case describes a well-known composition of Galilean and a scaling transformation for the NLS equation.

Next proposition extends the transformations above to the case of the Deformed RAKNS hierarchy.
Suppose  $\alpha_k(t),a,b\in\mathbb{R}$,  and
\begin{align*}
&X=ax+a\sum_{m\ge1}\binom{m+1}{1}(2b)^m\alpha_m(t),\\
&T_1=\left(\alpha_1(t)+\sum_{m\ge2}\binom{m+1}{2}(2b)^{m-1}\alpha_m(t)\right)a^2,\\
&T_2=\left(\alpha_2(t)+\sum_{m\ge3}\binom{m+1}{3}(2b)^{m-2}\alpha_m(t)\right)a^3,\\
&\quad\ldots
\end{align*}
\begin{thm}
Suppose that the function
\begin{equation*}
\Psi(x,\alpha_1(t),\alpha_2(t),\ldots)
\end{equation*}
is a solution of the deformed mixed RAKNS equation  \eqref{eq:tnls}.
Then the function
\begin{equation*}
\wtPsi(x,t)= a\Psi\left(X,T_1,T_2,\ldots\right)\exp\left\{-2ibx-i\sum_{m\ge1}(2b)^{m+1}\alpha_m(t)\right\}
\end{equation*}
is also the  solution of  \eqref{gnls}.
\end{thm}

\section{Some properties of the algebro-geometric solutions of the RAKNS hierarchy}

Here we give only extremely brief outline of the  well developed theory \cite{BBEIM}.

The theorems of the previous section follows immediately from the properties of the finite-gap solutions of RAKNS hierarchy but  they are valid also for all  possible solutions. For the first members of AKNS hierarchy this can be checked by straightforward calculation independent from the choice of the selected  class of solutions. Somehow for the higher order equation the direct proof is very involved and for the algebro-geometric solutions it works for any genus for any equation of the hierarchy.

For AKNS hierarchy  algebro-geometric solutions are parametrized by the moduli of the
hyperelliptic spectral curve $\Gamma =\{(w,\lambda)\}$ defined by the algebraic equation
\begin{equation}
w^2=\prod_{j=1}^{g+1}[(\lambda-\lambda_j)(\lambda-\lambda_j^\ast)], \quad \Im\lambda_j>0,
\end{equation}.
and have the following form
\begin{align}
&\psi(x,t_1,\ldots)= \frac{2K_0}{\rho}\frac{\Theta(\bZ)\Theta( \bU(x,t_1,\ldots)+\bZ-\bD)}%
{\Theta(\bZ-\bD)\Theta( \bU(x,t_1,\ldots)+\bZ)}\exp\{2 i \Phi(x,t_1,\ldots)\},\label{nls:p,q}\\
&\phi(x,t_1,\ldots)= 2\rho K_0\frac{\Theta(\bZ-\bD)\Theta( \bU(x,t_1,\ldots)+\bZ+\bD)}%
{\Theta(\bZ)\Theta( \bU(x,t_1,\ldots)+\bZ)}\exp\{-2 i \Phi(x,t_1,\ldots)\}.
\end{align}

Here  $\Theta$ is a multi-dimensional Riemann theta function defined on $\Gamma$ (see
\cite{BBEIM}),
\begin{equation*}
\bU(x,t_1,\ldots)=\bV^1 x+\sum_{j\ge1}\bV^{j+1} t_j,\quad
\Phi(x,t_1,\ldots)= -K_1 x-\sum_{j\ge1}K_{j+1} t_j.
\end{equation*}
Here above  $\bV^j$ are vectors of the $b$-periods, and  $K_j$ are coefficients of asymptotics of the normalized  Abelian integrals of the second kind  $\Omega_j$ (see \cite{BBEIM , Bake} for details):
\begin{align*}
&\oint_{a_m}d\Omega_j=0,\quad \oint_{b_m}d\Omega_j=(\bV^j)_m,\\
&\Omega_j(\Pp)= \mp i \left(2^{j-1}\lambda^j-K_j
+O\left(\lambda^{-1}\,\right)\right),
&&\Pp\to\Pp_{\infty}^{\pm}, \\
&\omega_0(\Pp)= \mp\left(\ln \lambda-\ln K_0
+O\left(\lambda^{-1}\,\right)\right),
&&\Pp\to\Pp_{\infty}^{\pm}, \\
& w = \pm\left( \lambda^{g+1}+O\left(\lambda^g\,\right)\right),
&&\Pp\to\Pp_{\infty}^{\pm}.
\end{align*}

The affine transformation of  $\Gamma$
\begin{equation}
\tau:\;(w,\lambda)\to(\wtw,\wtl),\quad \wtw=a^{g+1}w,\quad \wtl=a\lambda+b. \label{eq:transform}
\end{equation}
induces the related transformation of the abelian integrals:
\begin{align*}
&\wtO_j(\wtP)= \mp i \left(2^{j-1}\wtl^j-\wtK_j
+O\left(\wtl^{-1}\,\right)\right),
&&\wtP\to\wtP_{\infty}^{\pm}, \\
&\wto_0(\Pp)= \mp\left(\ln \wtl-\ln \wtK_0
+O\left(\wtl^{-1}\,\right)\right),
&&\wtP\to\wtP_{\infty}^{\pm}, \\
& \wtw = \pm\left( \wtl^{g+1}+O\left(\wtl^g\,\right)\right),
&&\wtP\to\wtP_{\infty}^{\pm}.
\end{align*}
Therefore,
\begin{align*}
&\wtO_j(\wtP)= \mp i \left(2^{j-1}a^j\lambda^j+\sum_{m=1}^j2^{j-1}\binom{j}{m}a^{j-m}\lambda^{j-m}b^m-\wtK_j
+O\left(\wtl^{-1}\,\right)\right),\\
&\wto_0(\wtP)=\mp\left(\ln \lambda+\ln a-\ln \wtK_0
+O\left(\wtl^{-1}\,\right)\right)
\end{align*}
and
\begin{align*}
&\wtV^{j}=a^j\bV^j+\sum_{m=1}^{j-1}2^m\binom{j}{m}a^{j-m}b^{m}\bV^{j-m},\\
&\wtK_j=K_j+2^{j-1}b^j,\\
&\wtK_0=a K_0.
\end{align*}

Let us consider the transformed solution
\begin{equation}  \label{eq:wtpsi}
\wtpsi(x,t_1,t_2,\ldots)=\frac{2\wtK_0}{\rho}\frac{\Theta(\bZ)\Theta( \wtU(x,t_1,\ldots)+\bZ-\bD)}%
{\Theta(\bZ-\bD)\Theta( \wtU(x,t_1,\ldots)+\bZ)}\exp\{2 i \wtPhi(x,t_1,\ldots)\},
\end{equation}
where
\begin{equation*}
\wtU(x,t_1,\ldots)=\wtV^1 x+\sum_{j\ge1}\wtV^{j+1} t_j,\quad
\wtPhi(x,t_1,\ldots)= -\wtK_1 x-\sum_{j\ge1}\wtK_{j+1} t_j.
\end{equation*}
Now it is easy to see that the solutions  \eqref{eq:wtpsi} and   \eqref{eq:nls} are connected by the relation,
\begin{equation}
\wtpsi(x,t_1,t_2,\ldots)= a\psi\left(X,T_1,T_2,\ldots\right)\exp\left\{-2ibx-i\sum_{m\ge1}(2b)^{m+1}t_m\right\}, \label{last}
\end{equation}
where $X $ and $T_j$ are defined by the formulas:
\begin{align*}
&X:=ax+a\sum_{m\ge1}\binom{m+1}{1}(2b)^mt_m,\\
&T_1:=a^2t_1+a^2\sum_{m\ge2}\binom{m+1}{2}(2b)^{m-1}t_m,\\
&T_2:=a^3t_2+a^3\sum_{m\ge3}\binom{m+1}{3}(2b)^{m-2}t_m,\\
&\quad\ldots
\end{align*}
Relation \eqref{last} completes the proof of the Theorem~1 for the algebro-geometric solutions. Theorems~2 and~3 can be proved in a similar way.

\section{Concluding remarks }
We established above very important symmetry properties of solutions of the individual RAKNS   hierarchy equations,  or for their mixed  and deformed  versions, (Theorems 1-3). This  was proved by us     for generic algebro-geometric solutions of any  finite genus. Therefore,  the same properties are also valid  for all degenerate cases of the algebro-geometric solutions,   including in particular
trigonometric, rational and quasi-rational solutions and a partially degenerate solutions as well.
 The direct calculation allows to prove the same statements,\footnote{ We checked it  for the orders $\leq 5$.}  for any
sufficiently  smooth solutions independently on their  nature. Therefore we believe that  the related
properties are absolutely universal.  It will be interesting to prove it  in whole generality  in a frame of Sato theory.

This work is supported by the Ministry of Education and Science of the Russian Federation (project 2527).



\begin{thebibliography}{10}

\bibitem{GNLS88}
M.~Lakshmanan, K.~Porsezian, and M.~Daniel, \emph{Effect of discreteness on the
  continuum limit of the {H}eisenberg spin chain}, Phys. Lett. A \textbf{133}
  (1988), no.~9, 483--488.

\bibitem{GNLS92}
K.~Porsezian, M.~Daniel, and M.~Lakshmanan, \emph{On the integrability aspects
  of the one-dimensional classical continuum isotropic {H}eisenberg spin
  chain}, J. Math. Phys. \textbf{33} (1992), 1807--1816.

\bibitem{GNLS93}
M.~Daniel, K.~Porsezian, and M.~Lakshmanan, \emph{On the integrable models of
  the higher order water wave equation}, Phys. Lett. A \textbf{174} (1993),
  no.~3, 237--240.

\bibitem{Ohta}
Y~Ohta, J~Satsuma, D.~Takahashi, and T.~Tokihiro, \emph{An elementary
  introduction to the {S}ato theory}, Prog. Theor. Phys. Suppl. \textbf{94}
  (1988), 210--241.

\bibitem{BBEIM}
E.~D. Belokolos, A.~I. Bobenko, V.~Z. Enol'skii, A.~R. Its, and V.~B. Matveev,
  \emph{Algebro-geometrical approach to nonlinear evolution equations},
  Springer Ser. Nonlinear Dynamics, Springer, 1994.

\bibitem{Hir73}
R.~Hirota, \emph{Exact envelope-soliton solutions of a nonlinear wave
  equation}, J. Math. Phys. \textbf{14} (1973), 805.

\bibitem{Hir06}
C.~Q Dai and J.~F. Zhang, \emph{New solitons for the {H}irota equation and
  generalized higher-order nonlinear {S}chr{\"o}dinger equation with variable
  coefficients}, J. Phys. A \textbf{39} (2006), 723--737.

\bibitem{AAS2}
A.~Ankiewicz, J.~M. Soto-Crespo, and N.~Akhmediev, \emph{Rogue waves and
  rational solutions of the Hirota equation}, Phys. Rev. E \textbf{81}
  (2010), 046602.

\bibitem{Hir13}
L.~Li, Zh. Wu, L.~Wang, and J.~He, \emph{High-order rogue waves for the Hirota
  equation}, Annals of Physics \textbf{334} (2013), 198--211.

\bibitem{HeLiP}
J.~S. He, Ch.~Zh. Li, and K.~Porsezian, \emph{Rogue waves of the Hirota and the
  Maxwell-Bloch equations}, Phys. Rev. E \textbf{87} (2013), no.~1, 012913.

\bibitem{GNLS13}
L.~H. Wang, K.~Porsezian, and J.~S. He, \emph{Breather and rogue wave solutions
  of a generalized nonlinear {S}chr{\"o}dinger equation}, Phys. Rev. E
  \textbf{87} (2013), no.~5, 053202.

\bibitem{AA14}
A.~Ankiewicz and N.~Akhmediev, \emph{High-order integrable evolution equation
  and its soliton solutions}, Phys. Lett. A \textbf{378} (2014), 358--361.

\bibitem{AKAC}
A.~Chowdury, D.~J. Kedziora, A.~Ankiewicz, and N.~Akhmediev, \emph{Breather
  solutions of the integrable quintic nonlinear {S}chr{\"o}dinger equation and
  their interactions}, Phys. Rev. E \textbf{91} (2015), 022919.

\bibitem{AKAC15}
D.~Kedziora,  A.~Ankiewicz,  A.~Chowdury  and N.~Akhmediev, \emph{Integrable equations of the infinite nonlinear Schr\"odingier equation hierarchy with time variable coefficients pp. 1-
12}
To appear  in Chaos (2015)

\bibitem{Bake}
H.~F. Baker, \emph{Abel's theorem and the allied theory including the theory of
  theta functions}, Cambridge, 1897.

\end{thebibliography}



\end{document}